\documentclass[aps,prb,a4paper,superscriptaddress,twocolumn,floatfix,showpacs]{revtex4}
\usepackage{graphicx}
\usepackage{latexsym}
\usepackage[]{amsmath}
\usepackage{amssymb}

\begin{document}
\newcommand{\chem}[1]{\ensuremath{\mathrm{#1}}}

\title{Muon-spin relaxation measurements on the dimerized \\ 
spin-$1/2$ chains \chem{NaTiSi_{2}O_{6}} and \chem{TiOCl}}

\author{P.\ J.\ Baker}
\affiliation{Clarendon Laboratory, University of Oxford, 
Parks Road, Oxford OX1 3PU, United Kingdom}

\author{S.\ J.\ Blundell}
\affiliation{Clarendon Laboratory, University of Oxford, 
Parks Road, Oxford OX1 3PU, United Kingdom}

\author{F.\ L.\ Pratt}
\affiliation{ISIS Muon Facility, ISIS, Chilton, Oxon., 
OX11 0QX, United Kingdom}

\author{T.\ Lancaster}
\affiliation{Clarendon Laboratory, University of Oxford, 
Parks Road, Oxford OX1 3PU, United Kingdom}

\author{M.\ L.\ Brooks}
\affiliation{Clarendon Laboratory, University of Oxford, 
Parks Road, Oxford OX1 3PU, United Kingdom}

\author{W.\ Hayes}
\affiliation{Clarendon Laboratory, University of Oxford, 
Parks Road, Oxford OX1 3PU, United Kingdom}

\author{M.\ Isobe}
\affiliation{Institute for Solid State Physics, University of Tokyo, 5-1-5 
Kashiwanoha, Kashiwa, Chiba 277-8581. Japan}

\author{Y.\ Ueda}
\affiliation{Institute for Solid State Physics, University of Tokyo, 5-1-5 
Kashiwanoha, Kashiwa, Chiba 277-8581. Japan}

\author{M.\ Hoinkis}
\affiliation{Experimentelle Physik 4, {Universit\"{a}t} {W\"{u}rzburg},
D-97074 {W\"{u}rzburg}, Germany}
\affiliation{Experimentalphysik II, {Universit\"{a}t} Augsburg, 
D-86159 Augsburg, Germany}

\author{M.\ Sing}
\affiliation{Experimentelle Physik 4, {Universit\"{a}t} {W\"{u}rzburg},
D-97074 {W\"{u}rzburg}, Germany}

\author{M.\ Klemm}
\affiliation{Experimentalphysik II, {Universit\"{a}t} Augsburg, 
D-86159 Augsburg, Germany}

\author{S.\ Horn}
\affiliation{Experimentalphysik II, {Universit\"{a}t} Augsburg, 
D-86159 Augsburg, Germany}

\author{R.\ Claessen}
\affiliation{Experimentelle Physik 4, {Universit\"{a}t} {W\"{u}rzburg},
D-97074 {W\"{u}rzburg}, Germany}

\date{\today}

\begin{abstract}
We report muon spin relaxation ($\mu$SR) 
and magnetic susceptibility investigations of two \chem{Ti^{3+}} 
chain compounds which each exhibit a spin gap at low temperature, 
\chem{NaTiSi_{2}O_{6}} and \chem{TiOCl}. 
From these we conclude that the spin gap in \chem{NaTiSi_{2}O_{6}} is 
temperature independent, with a value of $2\Delta=660(50)$~K,
arising from orbital ordering at $T_{\rm {OO}}=210$\,K; the associated 
structural fluctuations activate the muon spin relaxation rate up to 
temperatures above $270$~K.
In \chem{TiOCl} we find thermally activated spin fluctuations corresponding 
to a spin gap $2\Delta = 420(40)$~K below $T_{c1}=67$~K.
We also compare the methods used to extract the spin gap and the 
concentration of free spins within the samples from $\mu$SR and magnetic 
susceptibility data.
\end{abstract}

\maketitle

%\section{\label{sec:introduction} Introduction}
The interplay between spin, charge, and orbital degrees of freedom is 
particularly subtle in strongly-correlated oxides containing 
octahedrally-coordinated \chem{Ti^{3+}} ions ($t^{1}_{2g}$).
Such compounds are typically insulators because the $S=1/2$ spins are 
localized in the $t^{1}_{2g}$ orbitals.
In oxides containing chains of \chem{Ti^{3+}} ions, superexchange via 
oxygen gives rise to antiferromagnetic coupling.
A well known instability that can affect half-integer spin chains is the 
spin-Peierls (SP) transition~\cite{bray75},
in which magnetoelastic coupling dimerizes the chain and allows a spin-gap, 
$2\Delta$, to open below a characteristic temperature, $T_{\rm {SP}}$, 
resulting in a spin-singlet ground state.
(e.g. $T_{\rm SP} = 14$~K for \chem{CuGeO_3}~\cite{hase93}, $T_{\rm SP} = 
18$~K for \chem{MEM(TCNQ)_2}~\cite{huizinga79})
In contrast, for integer spins a Haldane gap will be present, precluding 
any low temperature instability~\cite{haldane83}. 

Recently, two oxides, \chem{NaTiSi_{2}O_{6}} (NTSO) and \chem{TiOCl}, have 
been intensively studied because they undergo dimerization transitions at 
unusually high temperatures. 
NTSO has the pyroxene structure with chains of \chem{TiO_6} octahedra that 
are only weakly coupled to one another~\cite{isobe02}.
TiOCl has \chem{TiO} bilayers within the $ab$ plane, well separated by 
\chem{Cl^{-}} ions. 
At low temperature dimerized chains of \chem{Ti^{3+}} ions form along the 
$b$ axis, with their spins coupled by direct exchange~\cite{shaz05}.
The magnetic susceptibility, $\chi$, drops sharply at 210~K for 
NTSO~\cite{isobe02} and 67~K for TiOCl~\cite{seidel03} due to the opening of 
a spin gap.
However, for both compounds, the ratio of the spin gap (which we will 
quote in Kelvin) to the dimerization 
temperature is larger than for the canonical SP case (for which 
$2\Delta/T_{\rm SP}=3.53$)~\cite{bray75}, demonstrating that the dimerization 
transitions are more complex than the canonical SP case, and pointing to 
the possible r\^{o}le of orbital physics.

X-ray diffraction studies of NTSO show that the \chem{Ti^{3+}} chains
dimerize below 210~K.~\cite{redhammer03}
Phonon anomalies measured using Raman scattering~\cite{konstantinovic04} 
are consistent with the dimerization being driven by an orbital ordering 
at $T_{\rm OO}=210$~K, below which the system is condensed in one of two 
possible orbitally ordered spin-singlet states, breaking translational 
symmetry.
This model is supported by a number of theoretical studies
\cite{hikihara04,streltsov06,vanwezel06}, although a composite $S=1$ Haldane 
chain ground state, with spin-triplet dimers of ferromagnetically coupled 
\chem{Ti^{3+}} spins, has also been proposed~\cite{popovic04}.
(See also Refs.~\onlinecite{streltsov06},~\onlinecite{popovic06}.)

Unusually, the dimerization of \chem{TiOCl} occurs in two stages.
There is a second order phase transition at $T_{c2}=91$~K into an 
incommensurate dimerized phase, and a first order transition at 
$T_{c1}=67$~K into a commensurate SP phase
\cite{seidel03,imai03,shaz05,hemberger05,ruckamp05}.
Although early work suggested that orbital fluctuations may play a 
dominant r\^{o}le in driving the transition~\cite{kataev03,sahadasgupta04,
lemmens04,hemberger05}, recent experiments have ruled this out~\cite{
ruckamp05,hoinkis05,zakharov06}.
In particular, optical measurements, in combination with a cluster 
calculation, revealed that the crystal field splitting is large enough to 
quench the orbital degree of freedom~\cite{ruckamp05}.
It appears that frustration between the two staggered chains in the bilayer 
crystal structure of TiOCl leads to an incommensurate SP state between 
$T_{c1}$ and $T_{c2}$~\cite{imai03,krimmel06,schonlieber06}.
The magnitude of the spin gap has previously been measured using 
optical techniques~\cite{lemmens04,caimi04} and NMR~\cite{imai03}.
The former suggest a value of $2\Delta \sim 430$~K, whereas the latter gave 
a value of $\Delta = 430 \pm 60$~K, which is very large compared 
to the observed transition temperatures. 

Muon spin relaxation ($\mu$SR) experiments~\cite{blundell99} probe 
magnetic ordering and dynamics from a microscopic viewpoint.
The applicability of this technique to spin-gapped systems has been 
demonstrated by studies of the canonical SP compounds 
\chem{CuGeO_3}~\cite{gm95,kojima97}, 
and \chem{MEM(TCNQ)_2} \cite{blundell97,lovett00}, and the Haldane chain 
compound \chem{Y_{2}BaNiO_{5}}~\cite{kojima95}.
In this paper we present the results of $\mu$SR experiments on NTSO and TiOCl.
We also made magnetic susceptibility measurements 
using a SQUID magnetometer
that allow us to 
compare our results with previous data~\cite{isobe02,seidel03,ruckamp05}.
We have measured the magnitude of the spin gap and found the concentration 
of unpaired spins in the dimerized state.

\begin{figure}
\includegraphics[width=\columnwidth]{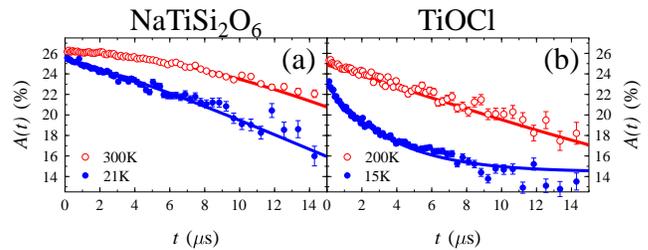}
\caption{\label{data} (Color online) 
Examples of muon decay asymmetry data in 
(a) \chem{NaTiSi_{2}O_{6}} and (b) \chem{TiOCl}, 
with fits to Eq.~\ref{fitfunc}.
}
\end{figure}

Our polycrystalline sample of \chem{NTSO} was prepared by a solid-state 
reaction~\cite{isobe02} of \chem{Na_{2}TiSi_{4}O_{11}}, \chem{Ti}, and 
\chem{TiO_{2}}.
The sample of \chem{TiOCl} was composed of small single crystals synthesized 
using standard vapour-transport techniques~\cite{schaefer58} from 
\chem{TiO_{2}} and \chem{TiCl_{3}}. 
Our $\mu$SR experiments were carried out using the MuSR and ARGUS 
spectrometers at the ISIS facility, United Kingdom.
These were done in zero applied magnetic field (ZF) and in small magnetic 
fields along the axis of the initial muon spin polarization.
Spin-polarized positive muons ($\mu^{+}$, mean lifetime $2.2\,\mu$s, momentum 
$28$\,MeV/$c$, $\gamma_{\mu}=2\pi \times 135.5$~MHzT$^{-1}$) were implanted 
into the polycrystalline samples where they stop within $\sim 1$\,ns.
The decay positron asymmetry function~\cite{blundell99}, $A(t)$, is 
proportional to the average spin polarization of the muons stopped within the 
sample.
Examples of the measured asymmetry spectra are presented in Fig.~\ref{data}. 

\begin{figure}
\includegraphics[width=\columnwidth]{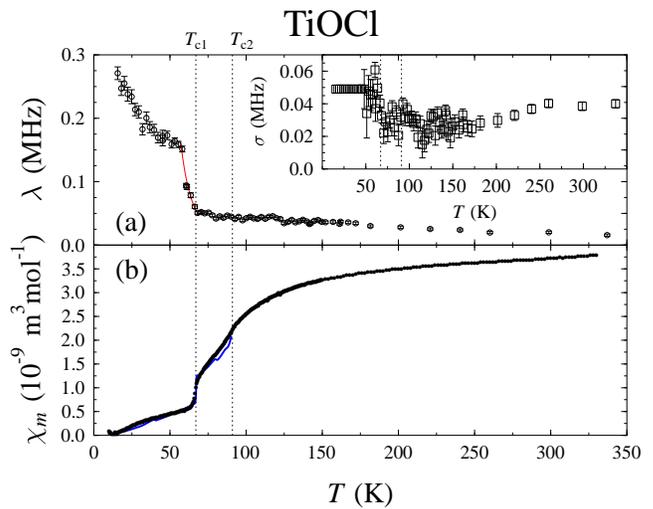}
\caption{\label{TiOCl}(Color online)  Fitted data for \chem{TiOCl}. 
(a) Electronic relaxation rate, $\lambda$ (Eq.~\ref{fitfunc}).
The red solid line is a fit to Eq.~\ref{satgap} between 58K and 68.2K, where 
the relaxation rate was found to be thermally activated. 
(Inset) Nuclear relaxation rate, $\sigma$, (Eq.~\ref{fitfunc}) showing a 
weak temperature dependence~\cite{tioclsigma}. 
(b) Magnetic susceptibility, $\chi_{m}$, after subtraction of the 
low-temperature Curie-Weiss tail.
The solid blue line is the fit, described in the text, using the values of 
the dimerization parameter from Ref.~\onlinecite{krimmel06} to define the 
temperature variation of the spin gap and applying Eq.~\ref{magsusc} to 
describe the susceptibility. 
}
\end{figure}

The absence of coherent muon precession, which would be 
indicative of a spontaneous magnetic field, together with the observation that
both high and low temperature spectra relax to the same value with 
negligible missing asymmetry (Fig.~\ref{data}), excludes the presence of 
long-range magnetic order in either material. 
The form of the muon spin relaxation is dependent on the distribution and 
time-dependence of the local magnetic fields around the site where the muon 
is implanted. 
In spin-gapped materials we write $A(t)$ as a product of 
relaxation functions~\cite{gm95,blundell97,lovett00}. 
We approximate a Gaussian Kubo-Toyabe function~\cite{blundell99} (which 
models the Gaussian distribution of randomly orientated nuclear spins) using 
$\exp[-(\sigma t)^2]$, use $\exp(-\lambda t)$ to describe the 
depolarization due to fluctuating electronic spins, and add a constant 
background $A_{\rm {BG}}$ to describe those muons landing outside the sample; 
giving a fitting function:
\begin{equation}
A(t) = A(0)\exp[-(\sigma t)^2]\exp(-\lambda t) + A_{\rm {BG}}.
\label{fitfunc}
\end{equation}
At high temperature the relaxation due to the fluctuations of nuclear spins 
dominates.
Because the observed value of $\sigma$ is small we do not see a recovery in 
the muon asymmetry at longer times; this is also due, in part, to the presence 
of the electronic fluctuations.
Fitting the data with $\sigma$ as a free parameter over the 
whole temperature range, $15-340$\,K, showed that in NTSO it was, within 
experimental error, temperature independent and it was subsequently fixed 
at $\sigma = 0.06$\,MHz. 
In \chem{TiOCl} the situation is more complicated~\cite{tioclsigma}. 
The values of $\sigma$ obtained for \chem{TiOCl} are shown in the inset to 
Fig.~\ref{TiOCl}(a).

\begin{figure}
\includegraphics[width=\columnwidth]{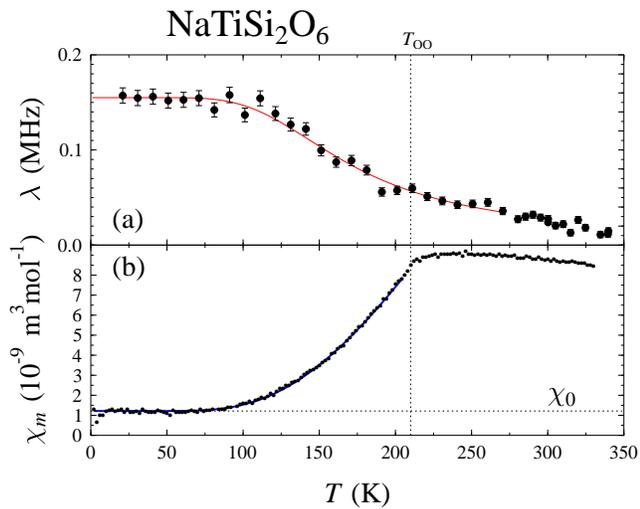}
\caption{\label{NTSO}(Color online) Fitted data for \chem{NaTiSi_{2}O_{6}}. 
(a) Electronic relaxation rate, $\lambda$ (Eq.~\ref{fitfunc}). The red 
solid line is a fit to Eq.~\ref{ffgap}.
(b) Magnetic susceptibility, $\chi_{m}$, after subtraction of the 
low-temperature Curie Weiss tail, with the temperature-independent 
paramagnetic response marked $\chi_0$. 
The blue line is a fit to Eq.~\ref{magsusc}.
}
\end{figure}

The relaxation of the muon spin due to the fluctuating electronic spins 
changes significantly as a function of temperature, dominating the 
relaxation at low temperature. 
Providing the electronic spin-fluctuation rate $\nu$ is fast compared with 
$\sqrt{\langle B^{2}_{\mu} \rangle} / \gamma_{\mu}$, where $B_{\mu}$ is the 
fluctuating field at the muon-site due to the electronic spins (the 
fast-fluctuation regime), then\cite{lovett00} $\lambda \propto 1 / \nu$.
The temperature variation of $\lambda$ is shown in Fig.~\ref{NTSO}(a) 
(\chem{NTSO}) and Fig.~\ref{TiOCl}(a) (\chem{TiOCl}).
At low temperature, one possible relaxation mechanism for the muon spin is 
via thermally activated electronic spins fluctuating across the spin gap
\cite{ehrenfreund77}.
This would lead to the relaxation rate $\lambda(T)$:
\begin{equation}
\lambda(T) = A\exp(2\Delta/T), \label{satgap}
\end{equation}
where $A$ is a constant.
X-ray measurements show that the degree of dimerization is essentially 
constant below $T_{\rm OO}$ (NTSO)~\cite{redhammer03} and $T_{c1}$ (TiOCl)
\cite{krimmel06}, so we take $\Delta$ as a constant for each compound. 
If an additional temperature-independent relaxation mechanism is present 
then:
\begin{equation}
\lambda(T) = \lambda_0 / [1+B\exp(-2\Delta/T)], \label{ffgap}
\end{equation}
where $B$ is a constant.

For NTSO, Eq.~\ref{ffgap} fits the data quite well at temperatures up to 
$T_{\rm OO}$ and even slightly above, suggesting that the relevant spin 
fluctuations in the paramagnetic state are controlled by a similar energy 
scale (in fact, the antiferromagnetic exchange 
$J\sim \Delta$)~\cite{streltsov06}. 
There is no sharp change in $\lambda$ at $T_{\rm {OO}}$, as might be 
expected if Haldane chains~\cite{popovic04} formed at that temperature, 
since the local field at the muon site should change significantly in that 
case.
The best fit to the data, found by including points up to 270~K, gives 
$2\Delta = 660 \pm 60$~K.

In contrast, for TiOCl, the behavior is more complex.
We see no significant changes in $\lambda$ either at $T_{c2}$ or at 
$T^{*}\simeq 120$~K, the temperature at which a pseudogap has been 
proposed~\cite{imai03,lemmens04,caimi04}.
However, $\lambda$ rises sharply below $T_{c1}$ and we are able to fit the 
$\lambda$ values between $58$~K and $68.2$~K to Eq.~\ref{satgap} yielding 
$2\Delta = 420 \pm 40$~K. 
This value is in agreement with the magnitude of the spin gap previously 
measured using other techniques~\cite{lemmens04,caimi04,imai03,NMRdetail}. 
At lower temperatures, $\lambda$ shows a much slower increase with decreasing 
temperature, and Eq.~\ref{fitfunc} only poorly describes the data.
Instead, a square root exponential, $\exp(-\sqrt{\Lambda t})$, as would be 
expected for a dilute distribution of slowly fluctuating electronic moments
\cite{uemura85}, provides a better description of the data.
This interpretation is supported by longitudinal-field measurements which 
imply that the fluctuations leave the fast fluctuation regime when 
$\lambda \gtrsim 0.16$~MHz (In contrast, similar measurements suggest that 
NTSO remains within the fast fluctuation regime at all measured temperatures).
At much lower temperatures, we speculate that the dilute distribution of 
electronic moments originates from unpaired spins due to defects and/or 
impurities.

%TiOCl figure

To gain a rough estimate of $n$, the concentration of unpaired spins within 
the samples, from the $\mu$SR data, we followed the method used in 
Ref.~\onlinecite{lovett00}.
We assume the unpaired spins to have $S=1/2$, given the spin of the 
\chem{Ti^{3+}} ions, although a defect in a chain would produce two free 
spins and any impurity creating the defect may itself have a spin, so our 
value will be larger than the concentration of defects and impurities.
Measurements were made in small magnetic fields, $\leq 5$mT for \chem{NTSO} 
and $2$~mT and $10$~mT for \chem{TiOCl}, applied along the 
direction of the initial muon spin polarization, at low temperature. 
The asymmetry spectra were fitted using the product of a longitudinal field 
Kubo-Toyabe function~\cite{blundell99} and an exponential relaxation to 
model the weak dynamics~\cite{lovett00}.
Fitting the data to this function we can extract the width of the distribution 
of local fields at the muon site. 
From this we estimate the concentration of unpaired spins surrounding the muon 
using the expression given in Ref.~\onlinecite{walstedt74} and adapted to the 
muon case in Ref.~\onlinecite{lovett00}.
For \chem{NTSO} this gives an impurity concentration $n=1.7(3)$\%, and 
for \chem{TiOCl} we find $n=1.1(2)$~\%.

Magnetic susceptibility, $\chi$, measurements (Fig.~\ref{NTSO}(b) and 
Fig.~\ref{TiOCl}(b)) showed that our samples were 
comparable with those used in previous studies. 
From the low temperature Curie tail in $\chi$ we estimate the 
concentration of unpaired spins to be $n=2.10(4)$~\% for \chem{NTSO} and 
$0.6(1)$~\% for \chem{TiOCl}. 
The close agreement between these values and those obtained from $\mu$SR 
supports our model describing the muon depolarization. 
The temperature-independent contribution to the susceptibility was 
$\chi_0 = 1.21(1) \times 10^{-9}\,{\mathrm m}^{3}{\mathrm {mol}}^{-1}$ for 
\chem{NTSO}, consistent with previous data~\cite{isobe02}.
For TiOCl we found that $\chi_0$ was negligible, once the contribution to 
the susceptibility from the dimerization was considered.
To extract the size of the spin gap from the magnetic susceptibility data, the 
Curie and temperature-independent terms were subtracted and we take the 
contribution, $\chi_m$, due to the thermal activation of spins across the 
spin gap of magnitude $2\Delta$ to be:
\begin{equation}
\chi_{m} = C \exp(- 2\Delta / k_{\mathrm B} T).
\label{magsusc}
\end{equation}
The result of fitting Eq.~\ref{magsusc} to the data for \chem{NTSO} is 
shown in Fig.~\ref{NTSO}(b).
Using a constant value for the spin gap in NTSO, this gives an excellent 
parameterization of the data from low temperature to $200$~K and leads to a 
value of the spin gap of $2\Delta = 595(7)$~K, somewhat smaller than the 
$\mu$SR measurements suggest.
The situation in \chem{TiOCl} is more complicated, since the dimerization 
varies strongly with temperature.
The structural dimerization gives alternating exchange constants along the 
chain $J_{1,2} = J(1 \pm \delta)$ and we take $\delta$ to vary linearly with 
the structural dimerization~\cite{cross79}.
To model the form of the susceptibility we used the intensity of the 
superstructure reflection observed in X-ray diffraction measurements
\cite{krimmel06}. 
For small values of $\delta$ the intensity of this peak varies as $\delta^{2}$.
The spin gap is proportional\cite{cross79} to $\delta^{2/3}$ and, using these 
results to convert the peak intensities into an effective spin gap, we 
then used Eq.~\ref{magsusc} to derive the line plotted in Fig.~\ref{TiOCl}(b). 
This was scaled by the spin gap value derived from the $\mu$SR measurements 
and the constant $C$ was chosen to fit the data, given we have no {\it a 
priori} values to choose for either case. 
A Curie-Weiss term was added to give the best fit to the low temperature data, 
a process complicated by oxygen adsorption, and so there remains a small 
discrepancy between the model and the data. 
Given that this model should only work in the limit $\Delta \gg T$, 
\cite{cross79} it can be seen to be remarkably successful in describing the 
form of the data below $80$~K, showing that the susceptibility is indeed 
varying with the structural dimerization. 
The breakdown of this model may explain why magnetic measurements suggest a 
lower value of $T_{c2}$ than the XRD measurements. 

\begin{table}[t]
\caption{\label{tab:gaps} 
Parameters derived from $\mu$SR and magnetic susceptibility data, 
$\chi$. The spin gap, $2\Delta$, the concentration of unpaired spins, $n$, 
and $T_{\rm {mf}} = 2\Delta / 3.53$, the temperature at which mean-field 
theory estimates that the SP transition should occur, given the value 
of $2\Delta$ measured by $\mu$SR.
}
\begin{ruledtabular}
\begin{tabular}{lccc}
Sample & \chem{NTSO} & \chem{TiOCl} \\
\hline
$2\Delta$~(K) $\mu$SR & 660(60) & 420(40) \\
$2\Delta$~(K) $\chi$ & 595(7) & - \\
$n$~(\%) $\mu$SR & 1.7(3) & 1.1(2) \\
$n$~(\%) $\chi$ & 2.10(4) & 0.6(1) \\ 
$T_{\rm {mf}}$~(K) & 190(20) & 120(10) \\
\end{tabular}
\end{ruledtabular}
\end{table}

%\section{\label{sec:conclusion} Conclusion}
In conclusion, we have measured the magnitude of the spin gap using $\mu$SR 
in the two \chem{Ti^{3+}} chain compounds \chem{NaTiSi_{2}O_{6}} and 
\chem{TiOCl}. 
Both the magnetic susceptibility and $\mu$SR data for \chem{NaTiSi_{2}O_{6}} 
are well described by assuming that the spin gap is independent of temperature 
below the transition at $T_{\rm {OO}} = 210$\,K, and we find the magnitude of 
the spin gap, $2\Delta = 660 \pm 60$\,K, to be consistent with magnetic 
susceptibility measurements (Table~\ref{tab:gaps}) and recent theoretical work
\cite{isobe02,hikihara04,streltsov06}.
\chem{TiOCl} shows a thermally activated increase in the muon relaxation rate 
below $T_{c1} = 67$~K.
Our value of $2\Delta = 420 \pm 40$\,K is consistent with previous 
NMR data~\cite{imai03,NMRdetail} and optical measurements
\cite{caimi04,lemmens04}. 
Using mean-field theory to compare the measured spin gap in \chem{TiOCl} to 
the observed transitions, we note (Table~\ref{tab:gaps}) that 
$T_{\rm {mf}} \simeq T^{*}$, the temperature where psuedogap formation has 
been observed~\cite{imai03,caimi04,lemmens04}, suggesting that $T^{*}$ is 
analogous to $T_{\rm sp}$ in \chem{TiOCl}, yet the system dimerizes at a far 
lower temperature. 
It is likely that the frustration between chains in the bilayer structure 
lowers the temperature at which dimerization occurs.
In \chem{TiOCl} we can directly relate previously reported superstructure 
reflections~\cite{krimmel06} to the magnetic susceptibility.
Comparing the concentrations of unpaired spins determined by $\mu$SR and 
magnetic susceptibility measurements (Table~\ref{tab:gaps}) we find that 
for both compounds the two methods are consistent, although our assumptions 
suggest that both techniques overestimate the impurity concentration.

We are grateful to P.\ J.\ C.\ King and the staff of the ISIS Pulsed Muon 
facility for experimental assistance.


\begin{thebibliography}{**}
\bibitem{tioclsigma}
While there was no trend in the temperature dependence, there 
were significant fluctuations in the value of $\sigma$ measured in 
\chem{TiOCl}, and so it was left as a free parameter whilst fitting the data.
We include data from two separate experiments and the slight change in the 
relaxing background strongly affected $\sigma$, despite the fact that the 
values of $\lambda$ were in excellent agreement.
\bibitem{NMRdetail}
The model of Ref.~\onlinecite{ehrenfreund77}, describing 
NMR data, implies that the $\Delta$ value measured in Ref.~\onlinecite{imai03} 
corresponds to $2\Delta$ in the conventional notation.
\end{thebibliography}
\end{document}